\documentclass{PoS_arXiv}

\usepackage{amsmath,amssymb,amsfonts,cite}

\input paperdef
\newcommand{\decayH}{\Stopz \to \Sbote H^+}
\newcommand{\SE}{S1}
\newcommand{\SZ}{S2}
\graphicspath{{figs/}}

\title{Consistent Higher-Order Corrections to 
\boldmath{$\Stopi \to \Sbotj H^+$} in the Complex MSSM}

\ShortTitle{\boldmath{$\Stopi \to \Sbotj H^+$} in the cMSSM}

\author{\speaker{Sven Heinemeyer}\\
        Instituto de F\'isica de Cantabria (CSIC-UC), Santander, Spain\\
        E-mail: \email{Sven.Heinemeyer@cern.ch}}

\author{Heidi Rzehak\\
        Institut f\"ur Theoretische Physik, 
        Karlsruhe Institute of Technology, D--76128 Karlsruhe, Germany\\
        Albert-Ludwigs-Universitaet Freiburg, Physikalisches Institut, D--79104
        Freiburg, Germany\\
        E-mail: \email{heidi.rzehak@physik.uni-freiburg.de}}

\author{Christian Schappacher\\
        Institut f\"ur Theoretische Physik, 
        Karlsruhe Institute of Technology, D--76128 Karlsruhe, Germany\\
        E-mail: \email{cs@particle.uni-karlsruhe.de}}

\abstract{
We review an analysis of a consistent renormalization of the top and bottom
quark/squark sector of the MSSM with complex parameters
(cMSSM). 
Various renormalization schemes are defined, analyzed analytically and
tested numerically in the decays $\Stopz \to \Sboti\, H^+/W^+$ ($i = 1,2$).
No scheme is found that produces numerically acceptable results
over all the cMSSM parameter space, where problems occur mostly already
for real parameters.
Some numerical examples for $\Ga(\Stopz \to \Sbote H^+)$ in our
preferred scheme, ``$\mb,\,\Ab$~\DRbar'' are shown.
}

\FullConference{Third International Workshop on Prospects for Charged Higgs Discovery at Colliders - CHARGED2010,\\
		September 27-30, 2010\\
		Uppsala Sweden}

\begin{document}


\section{Introduction}

One of the main tasks of the LHC is to search for Supersymmetry
(SUSY)~\cite{mssm}. 
The Minimal Supersymmetric Standard Model (MSSM) predicts two scalar
partners for all Standard Model (SM) fermions as well as fermionic
partners to all SM bosons. 
Of particular interest are the scalar partners of the heavy SM
quarks, the scalar top quarks, $\Stopi$ ($i = 1,2$) and scalar bottom
quarks $\Sbotj$ ($j = 1,2$) due to their large Yukawa couplings. 
Depending on the SUSY mass patterns, possibly important decay modes of
the scalar tops are,
\begin{align}
\label{stsbH}
&\Stopi \to \Sbotj H^+ \quad (i,j = 1,2)~, \\
\label{stsbW}
&\Stopi \to \Sbotj W^+ \quad (i,j = 1,2)~,
\end{align}
where $H^+$ denotes the (positively) charged MSSM Higgs boson.
These processes can constitute a large part of
the total stop decay width, and, in case of decays to a Higgs boson, they
can serve as a source of charged Higgs bosons in cascade decays at the LHC.
                      
For a precise prediction of the partial decay widths corresponding to
\refeq{stsbH} and \refeq{stsbW}, at least the one-loop level contributions
have to be taken into account.
This in turn requires a renormalization of the relevant sectors,
especially a simultaneous renormalization of the top and bottom
quark/squark sector. 
Due to the $SU(2)_L$ invariance of the left-handed scalar top and
bottom quarks, these two sectors cannot be treated independently.
Within the framework of the MSSM with complex parameters (cMSSM) 
we review the analysis of various bottom quark/squark sector renormalization
schemes~\cite{SbotRen}, while for the top quark/squark sector
a commonly used on-shell renormalization scheme is applied throughout
all the investigations. An extensive list of earlier analyses and
corresponding references can be found in \citere{SbotRen}.  
The evaluation of the partial decay widths of the scalar top quarks are
being implemented into the Fortran code 
{\tt FeynHiggs}~\cite{feynhiggs,mhiggslong,mhiggsAEC,mhcMSSMlong}. 


\section{The bottom/sbottom sector and its renormalization}

\subsection{The generic structure}
\label{sec:generic}

The bilinear part of the  Lagrangian with top and bottom squark fields,
$\Stop$ and $\Sbot$, 
\begin{align}
\cL_{\Stop/\Sbot\text{ mass}} &= - \begin{pmatrix}
{{\tilde{t}}_{L}}^{\dagger}, {{\tilde{t}}_{R}}^{\dagger} \end{pmatrix}
\matr{M}_{\tilde{t}}\begin{pmatrix}{\tilde{t}}_{L}\\{\tilde{t}}_{R}
\end{pmatrix} - \begin{pmatrix} {{\tilde{b}}_{L}}^{\dagger},
{{\tilde{b}}_{R}}^{\dagger} \end{pmatrix}
\matr{M}_{\tilde{b}}\begin{pmatrix}{\tilde{b}}_{L}\\{\tilde{b}}_{R} 
\end{pmatrix}~,
\end{align}
contains the stop and sbottom mass matrices
$\matr{M}_{\tilde{t}}$ and $\matr{M}_{\tilde{b}}$,
given by 
\begin{align}\label{Sfermionmassenmatrix}
\matr{M}_{\tilde{q}} &= \begin{pmatrix}  
 M_{\tilde Q_L}^2 + m_q^2 + M_Z^2 c_{2 \beta} (T_q^3 - Q_q \sw^2) & 
 m_q \Xq^* \\[.2em]
 m_q \Xq &
 M_{\tilde{q}_R}^2 + m_q^2 +M_Z^2 c_{2 \beta} Q_q \sw^2
\end{pmatrix} 
\end{align}
with
$\Xq = \Aq - \mu^*\kappa$ and $\kappa = \{\cot\beta, \tan\beta\}$ for
$q = \{t, b\}$. 
$M_{\tilde Q_L}^2$ and $M_{\tilde{q}_R}^2$ are the soft SUSY-breaking mass
parameters. $m_q$ is the mass of the corresponding quark.
$Q_{{q}}$ and $T_q^3$ denote the charge and the isospin of $q$, and
$A_q$ is the trilinear soft SUSY-breaking parameter.
The mass matrix can be diagonalized with the help of a unitary
 transformation ${\matr{U}}_{\tilde{q}}$, 
\begin{align}\label{transformationkompl}
\matr{D}_{\tilde{q}} &= 
\matr{U}_{\tilde{q}}\, \matr{M}_{\tilde{q}} \, 
{\matr{U}}_{\tilde{q}}^\dagger = 
\begin{pmatrix} \msqe^2 & 0 \\ 0 & \msqz^2 \end{pmatrix}~, \qquad
{\matr{U}}_{\tilde{q}}= 
\begin{pmatrix} U_{\tilde{q}_{11}}  & U_{\tilde{q}_{12}} \\  
                U_{\tilde{q}_{21}} & U_{\tilde{q}_{22}}  \end{pmatrix}~. 
\end{align}
The scalar quark masses, $\msqe$ and $\msqz$, will always be mass
ordered, i.e.\  
$m_{\tilde{q}_1} \le m_{\tilde{q}_2}$:
\begin{align}
m_{\tilde{q}_{1,2}}^2 &= \edz \KL M_{\tilde{Q}_L}^2 + M_{\tilde{q}_R}^2 \KR 
       + m_q^2 + \edz T_q^3 \MZ^2 c_{2\be} \non \\
&\quad \mp \edz \sqrt{\KKL M_{\tilde{Q}_L}^2 - M_{{\tilde{q}_R}}^2 
       + \MZ^2 c_{2\be} (T_q^3 - 2 Q_q \sw^2) \KKR^2 + 4 m_q^2 |\Xq|^2}~.
\label{MSbot}
\end{align}


\subsection{Renormalization of the bottom/sbottom sector}

The field renormalization constants of the bottom/sbottom (as well as of
the top/stop) sector are chosen according to an on-shell
prescription~\cite{SbotRen}. 

The parameter renormalization can be performed as follows, 
\begin{align}
\matr{M}_{\sq} &\to \matr{M}_{\sq} + \de\matr{M}_{\sq}
\end{align}
which means that the parameters in the mass matrix $\matr{M}_{\sq}$ 
are replaced by the renormalized parameters and a counterterm. After the
expansion $\de\matr{M}_{\sq}$ contains the counterterm part,
\begin{align}\label{proc1a}
\de\matr{M}_{\sq_{11}} &= \de M_{\tilde Q_L}^2 + 2 m_q \de m_q 
- M_Z^2 c_{2 \beta}\, Q_q \, \de \sw^2 + (T_q^3 - Q_q \sw^2) 
  ( c_{2 \beta}\, \de M_Z^2 + M_Z^2\, \de c_{2\beta})~, \\\label{proc1b}
\de\matr{M}_{\sq_{12}} &= (\Aq^*  - \mu \kappa)\, \de m_q 
+ m_q (\de \Aq^* - \mu\, \de \kappa - \kappa \, \de \mu)~, \\\label{proc1c}
\de\matr{M}_{\sq_{21}} &=\de\matr{M}_{\sq_{12}}^*~, \\\label{proc1d}
\de\matr{M}_{\sq_{22}} &= \de M_{\tilde{q}_R}^2 
+ 2 m_q \de m_q +  M_Z^2 c_{2 \beta}\, Q_q \, \de \sw^2
+ Q_q \sw^2 ( c_{2 \beta}\, \de M_Z^2+ M_Z^2\, \de c_{2 \beta})~.
\end{align}

Another possibility for the parameter renormalization is to start out
with the physical parameters which corresponds to
the replacement:
\begin{align} \label{proc2}
\matr{U}_{\tilde{q}}\, \matr{M}_{\tilde{q}} \, 
{\matr{U}}_{\tilde{q}}^\dagger &\to\matr{U}_{\tilde{q}}\, \matr{M}_{\tilde{q}} \, 
{\matr{U}}_{\tilde{q}}^\dagger + \matr{U}_{\tilde{q}}\, \de \matr{M}_{\tilde{q}} \, 
{\matr{U}}_{\tilde{q}}^\dagger =
\begin{pmatrix} \msqe^2 & Y_q \\ Y_q^* & \msqz^2 \end{pmatrix} +
\begin{pmatrix}
\de \msqe^2 & \de Y_q \\ \de Y_q^* & \de \msqz^2
\end{pmatrix}~,
\end{align}
where $\de \msqe^2$ and  $\de \msqz^2$ are the counterterms 
 of the squark masses squared. $\de Y_q$ is the
 counter\-term\footnote{The unitary 
     matrix $\matr{U}_{\tilde{q}}$ can be expressed by a mixing angle
     $\theta_{\tilde{q}}$ and
     a corresponding phase $\varphi_{\tilde{q}}$. Then the
     counterterm  $\de Y_q$ can be related to the counterterms of the
     mixing angle and the phase (see \citere{mhcMSSM2L}).}   to the squark
 mixing parameter $Y_q$ (which vanishes
   at tree level, $Y_q = 0$, and corresponds to the 
 off-diagonal entries in $\matr{D}_{\tilde{q}} =\matr{U}_{\tilde{q}}\,
 \matr{M}_{\tilde{q}} \,  
{\matr{U}}_{\tilde{q}}^\dagger$, see~\refeq{transformationkompl}). Using
\refeq{proc2} 
 one can express $\de\matr{M}_{\sq}$ by the counterterms $\de \msqe^2$,
 $\de \msqz^2$ and $\de Y_q$. Especially for $\de\matr{M}_{\sq_{12}}$
 one yields
\begin{align}\label{dMsq12physpar}
\de\matr{M}_{{\sq}_{12}} &=
U^*_{\sq_{11}} U_{\sq_{12}}
(\de \msqe^2 - \de \msqz^2) +
U^*_{\sq_{11}} U_{\sq_{22}} \de Y_q + U_{\sq_{12}}
U^*_{\sq_{21}} \de Y_q^*~.
\end{align}
For the top/stop sector we use an on-shell renormalization, see e.g.\
\citeres{dissHR,mhcMSSM2L,SbotRen}.
The various options to renormalize the bottom/sbottom sector are listed
in \refta{tab:RS}.

\begin{table}[ht!]
\renewcommand{\arraystretch}{1.5}
\BC
\begin{tabular}{|c||c|c|c|c||c|}
\hline
scheme & $m_{\tilde b_{1,2}}$ & $\mb$ & $\Ab$ & $Y_b$ & name 
                                                           \\ \hline\hline
{\small analogous to the $t/\Stop$ sector:} 
 ``OS'' & OS & OS & & OS  
& RS1 \\ \hline
``$\mb,\,\Ab$~\DRbar'' & OS & \DRbar\ & \DRbar\ & 
& RS2 \\ \hline
``$\mb,\, Y_b$~\DRbar'' & OS & \DRbar\ & & \DRbar\ 
& RS3 \\ \hline
``$\mb$~\DRbar, $Y_b$~OS'' & OS & \DRbar\ & & OS
& RS4 \\ \hline
``$\Ab$~\DRbar, $\re Y_b$~OS'' & OS & & \DRbar & $\re Y_b$:\, OS
& RS5 \\ \hline
``$\Ab$~vertex, $\re Y_b$~OS'' & OS & & vertex & $\re Y_b$:\, OS
& RS6 \\ \hline
\end{tabular}
\caption{Summary of the six renormalization schemes for the
  $b/\Sbot$~sector investigated in \citere{SbotRen}. 
  Blank entries indicate dependent
  quantities. $\re Y_b$ denotes that only the real part of 
  $Y_b$ is renormalized on-shell, while the imaginary part is a
  dependent parameter.}
\label{tab:RS}
\renewcommand{\arraystretch}{1.0}
\EC
\end{table}


\subsection{Summary of the renormalization scheme analysis}
\label{sec:scheme-ana}

A bottom quark/squark sector
renormalization scheme always contains dependent counterterms which can
be expressed by the independent ones.
According to our six definitions, these dependent parameters can be
$\de\mb$, $\de\Ab$ or $\de Y_b$. 
A problem can occur when the MSSM parameters are chosen such that the
independent counterterms (nearly) drop out of the relation determining
the dependent counterterms. This can lead to (unphysically) large
counterterm contributions in such a case.
As it was shown in \citere{SbotRen} it is
possible already in very generic SUSY scenarios to find a set of MSSM
parameters which show this behaviour for each of the chosen
renormalization schemes.
Consequently, it appears to be difficult {\em by construction} to
define a 
renormalization scheme for the bottom quark/squark sector (once
the top quark/squark
sector has been defined) that behaves well for the full MSSM parameter
space. One possible exception could be a pure \DRbar\ scheme, which,
however, is not well suited for processes with external top
squarks and/or bottom squarks. 

The analytical and numerical analysis performed in \citere{SbotRen}
identfied RS2 as ``preferred scheme''. This schemes showed the
``relatively most stable'' behavior, 
problems only occur for maximal sbottom mixing, 
$|U_{\Sbot_{11}}| = |U_{\Sbot_{12}}|$, where a divergence in $\de Y_b$
appears. 
On the other hand,
other schemes with $\de\mb$ or $\de\Ab$ as dependent counterterms 
generally exhibit problems in larger parts of the parameter
MSSM space and may induce large effects, since $\mb$ (or the bottom Yukawa 
coupling) and $\Ab$ enter prominently into the various couplings of the
Higgs bosons to other particles.


\section{Numerical Example}

In this section we show some example results for
$\Ga(\decayH)$~\cite{SbotRen}. This decay mode can
serve potentially as a source of charged MSSM Higgs bosons in SUSY
cascade decays. The parameters are chosen according to the two scenarios
\SE\ and \SZ\ as defined in \refta{tab:para}.

\begin{table}[t!]
\renewcommand{\arraystretch}{1.5}
\BC
\begin{tabular}{|c||r|r|r|r|r|r|r|r|}
\hline
Scen.\ & $\MHp$ & $\mstz$ & $\mu$ & $\At$ & $\Ab$ & $M_1$ & $M_2$ & $M_3$ 
\\ \hline\hline
\SE & 150 & 600 & 200 &  900 &  400 & 200 & 300 & 800 
\\ \hline
\SZ & 180 & 900 & 300 & 1800 & 1600 & 150 & 200 & 400  
\\ \hline
\end{tabular}
\caption{MSSM parameters for the initial numerical investigation; all
parameters are in GeV. 
We always set $\mb^{\MSbar}(\mb) = 4.2 \gev$.
In our analysis we use 
$M_{\tilde Q_L}(\Stop) = M_{\Stop_R} = M_{\Sbot_R} =: \msusy$, where $\msusy$
  is chosen such that the above value of $\mstz$ is realized.
The parameters entering the scalar lepton sector and/or the first two
generations do not play a relevant role in our analysis.
The values for $\At$ and $\Ab$ are chosen such that charge- or
color-breaking minima are avoided.
}
\label{tab:para}
\EC
\renewcommand{\arraystretch}{1.0}
\end{table}

\begin{figure}[htb!]
\begin{center}
\includegraphics[width=0.49\textwidth,height=5.5cm]{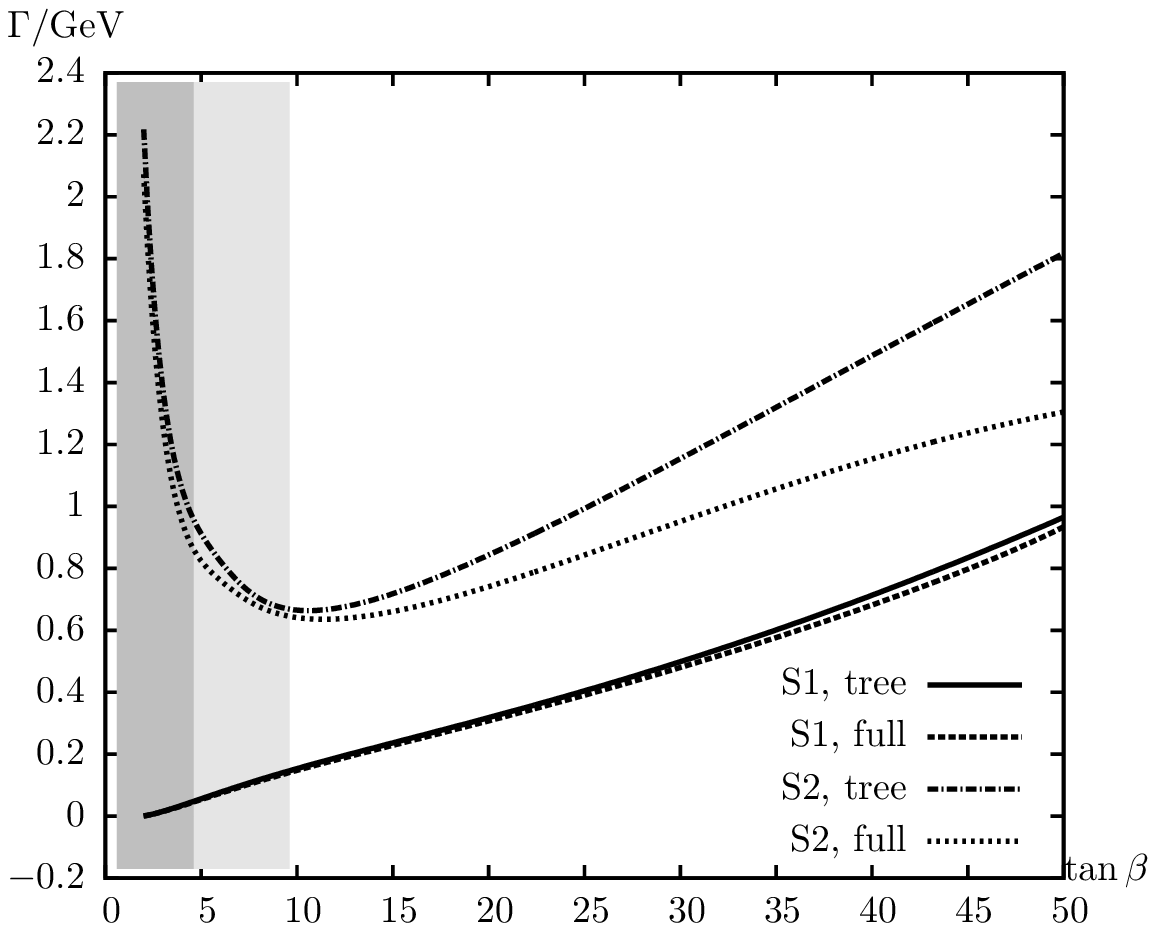}
\hspace{-2mm}
\includegraphics[width=0.49\textwidth,height=5.5cm]{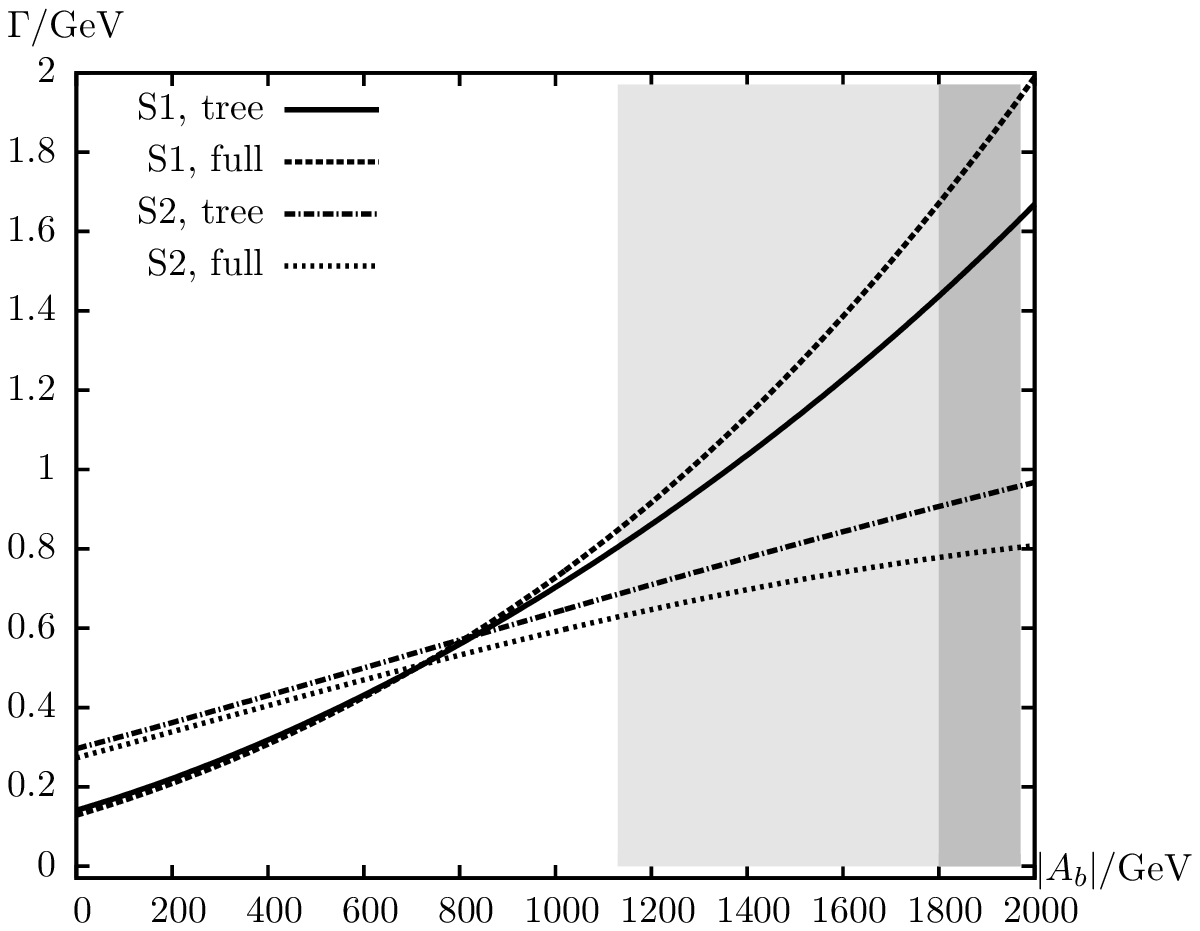} 
\\[1em]
\includegraphics[width=0.49\textwidth,height=5.5cm]{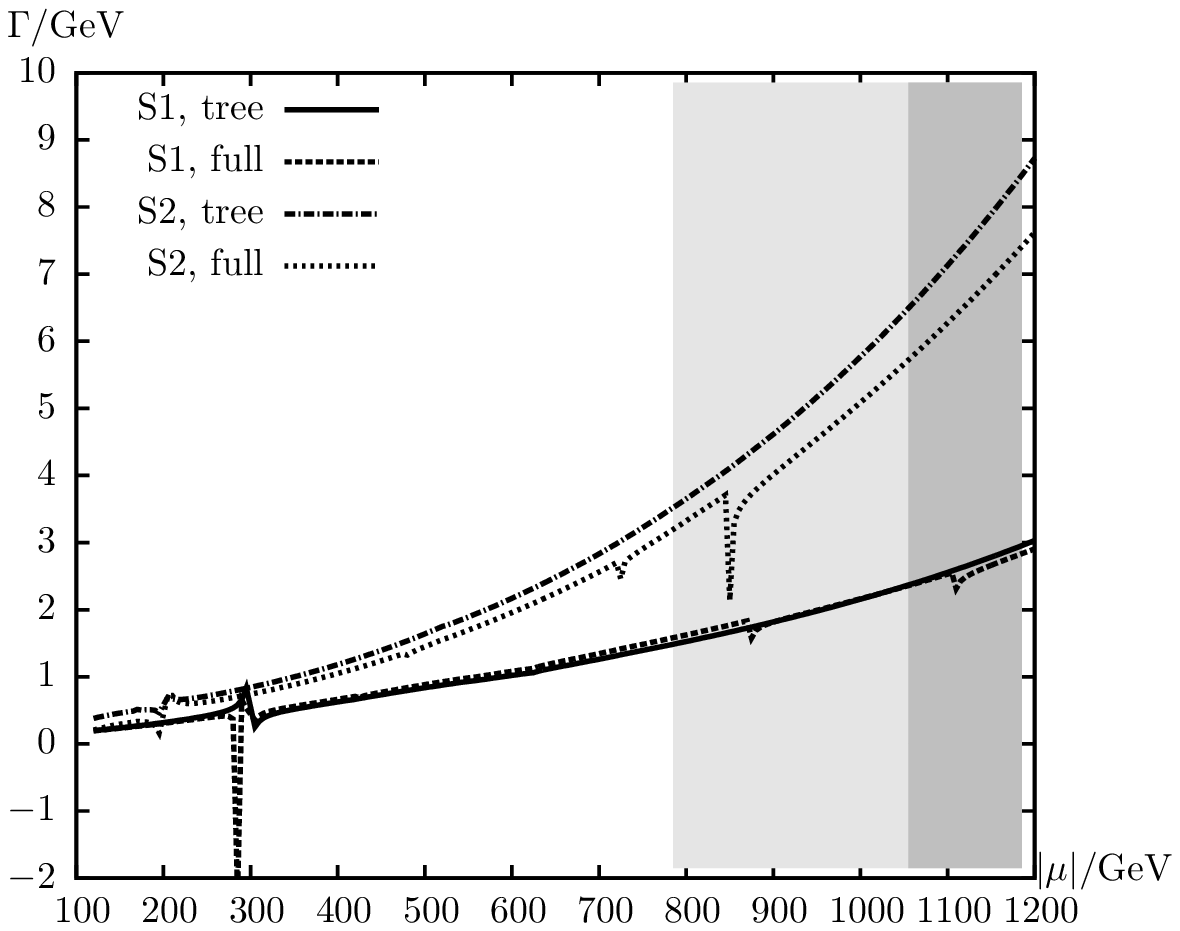}
\hspace{-2mm}
\includegraphics[width=0.49\textwidth,height=5.5cm]{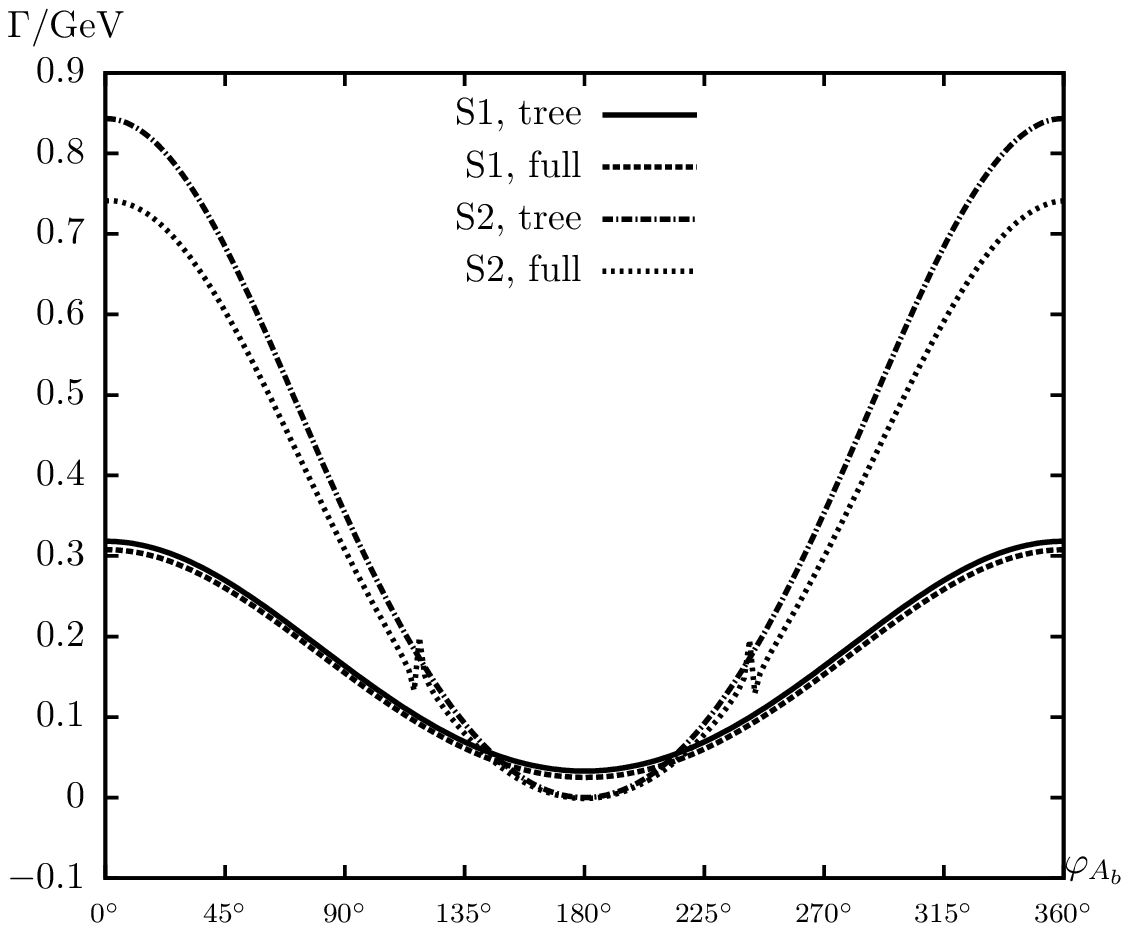}
\vspace{2em}
\caption{
$\Ga(\decayH)$. Tree-level and full one-loop corrected partial decay widths 
  for the renormalization scheme RS2. The parameters are chosen according to
  the scenarios \SE\ and \SZ. 
  For \SE\ the grey region is excluded and for \SZ\ the dark grey region 
  is excluded.
  Upper left plot: $\tb$ varied.
  Upper right plot: $\tb = 20$ and $|\Ab|$ varied.
  Lower left plot: $\tb = 20$ and $|\mu|$ varied.
  Lower right plot: $\tb = 20$ and $\varphi_{\Ab}$ varied.}
\label{fig:st2sb1H}
\end{center}
\end{figure}

In \reffi{fig:st2sb1H} we show the partial decay width 
$\Ga(\decayH)$ as a function of $\tb$ (upper left), as a function of
$\Ab$ (upper right), as a function of $\mu$ (lower left) and as a
function of $\phiab$ (lower right plot). ``tree'' denotes the tree-level value
and ``full'' is the decay width including all one-loop 
corrections (including hard QED and QCD radiation, see \citere{SbotRen}
for details)%
\footnote{
Corrections from imaginary parts of external leg self-energy
contributions~\cite{imim} are not included.}%
.~For \SE\ the grey region is excluded and for \SZ\ the dark grey region 
is excluded.
The spikes and dips visible in the lower left plot are due to various
particle thresholds, while the first dip in \SE\ is due to 
$|U_{\Sbot_{11}}| \approx |U_{\Sbot_{12}}|$.
The two spikes in the lower right plot are also due to 
$|U_{\Sbot_{11}}| \approx |U_{\Sbot_{12}}|$,
which leads to a divergence
in RS2, which, however, is confined to very narrow
intervals.
The loop corrections, as can be observed in all four plots, are
relatively modest, staying below $\sim 25\%$ for all parameters. The
fact of relatively small one-loop corrections shows that no unphysically
large contributions via large counterterms are introduced, a
characteristic of a suitable renormalization scheme. 

The real quantity of interest at the LHC is the $\br(\decayH)$. This,
however, requires the evaluation of {\em all} decay modes (at the same
level of accuracy). The corresponding results will be presented
elsewhere~\cite{Stop2decay}. 


\subsection*{Acknowledgements}

S.H.\ thanks the organizers of {\em cHarged 2010} for the invitation and
an inspiring workshop. 
Work supported in part by the European Community's Marie-Curie Research
Training Network under contract MRTN-CT-2006-035505
`Tools and Precision Calculations for Physics Discoveries at Colliders'.



\end{document}